\documentclass[11pt,twoside]{article}
\usepackage{CAGN2016}
\usepackage{graphicx}

\usepackage[T1]{fontenc} 

\usepackage{latexsym}
\usepackage{verbatim}
\usepackage[utf8]{inputenc}
\usepackage[T1]{fontenc}

\begin{document}

\vskip 1.0cm \markboth{H. Monteiro et al.}{physical and chemical
  conditions of NGC~6778} \pagestyle{myheadings}
%
%
\vspace*{0.5cm}
\parindent 0pt{Poster}


\vspace*{0.5cm} \title{Investigating spatial variation of the physical
  and chemical conditions of NGC~6778.}

\author{H. Monteiro$^1$, Garcia-Rojas$^{2,4}$, J., Jones$^{2,4}$, D.,
  Corradi$^{2,3,4}$, R.\\
  and Rodriguez-Gil$^{2,4}$, P. }
\affil{  $^1$Universidade Federal de Itajub\'a - UNIFEI\\
  $^2$Instituto de Astrofísica de Canarias (IAC), Vía Láctea s/n, E38200, La Laguna, Tenerife, Spain\\
  $^3$GRANTECAN, Cuesta de San José s/n, E-38712 , Breña Baja, La
  Palma, Spain\\
  $^4$Departamento de Astrofísica, Universidad de La Laguna, La
  Laguna, E-38206, Tenerife, Spain }

\begin{abstract}
  A number of planetary nebulae show binary central stars and
  significant abundance discrepancies between values estimated from
  colisionally excited lines when compared to the same abundances
  estimated from recombination lines. One approach to investigate this
  yet unsolved problem is using spatially resolved images of emission
  lines in an attempt to detect a possibly distinct metal rich
  component in the nebula. In this work we present results of
  spatially resolved abundance analysis of NGC 6778 based on data
  gathered from VLT VIMOS-IFU. We discuss the spatial variations found
  as well as possible limitations of the method in answering questions
  about abundance variations.

\end{abstract}

\section{Introduction}

Chemical abundances in planetary nebulae are important as a tool to
study stellar evolution, estimate effects of internal stellar
nucleosyntesis and mixing, among others issues related to chemical
evolution. Because of that PNe are used to study the chemical
evolution in the Galaxy and in other nearby galaxies.

It is well known (see e.g. Osterbrock \& Ferland 2006) that in
photoionized nebulae – both H II regions and planetary nebulae –
optical recombination lines (ORLs) provide chemical abundance values
that are systematically larger than those obtained using collisionally
excited lines (CELs). The abundance discrepancy factor (adf) between
ORLs and CELs is usually between 1.5 and 3 (see e.g. García–Rojas \&
Esteban 2007, Liu 2012), but in planetary nebulae (PNe) it has a
significant tail extending to much larger values.  This is generally
known as the abundance discrepancy problem. It has been around for
more than seventy years (Wyse 1942), and is one of the major
unresolved problems in nebular astrophysics. The problem has
far-reaching consequences on the measurement of abundances throughout
the Universe, as the chemical content of near and faraway galaxies is
most often done using CELs from their ionized gas (e.g. Hamann et
al. 2002).

Corradi et al. (2015) have recently shown that the largest abundance
discrepancies are reached in planetary nebulae with close binary
central stars. For instance, in the PNe Abell 46, Ou5, and NGC 6778,
which have binary central stars with orbital periods of a few hours,
they found $O^{2+}/H^+$ adfs larger than 40, and as high as 300 in the
inner regions of Abell 46.  Spectroscopic analysis supports the
previous interpretation that two different gas phases coexist in these
nebulae (e.g. Liu et al. 2006, Tsamis et al. 2008): hot gas at $10^4$
K with standard metallicity where the CELs can be efficiently excited,
and a much cooler ( $10^3$ K) plasma with a highly enhanced content of
heavy elements (which is likely the cause of the cooling) where only
ORLs form. How much each gas component contributes to the total mass,
and how they are distributed and mixed, is basically unknown.

In this work we show the results for the object NGC 6778 obtained from
VLT IFU data and spatially resolved abundance analysis. The present
effort is part of a larger project that aims to understand the large
discrepancy in PNe.

\section{Observational data}

The observations were obtained with the instrument VIMOS-IFU, attached
to ESO-VLT-U3 on the 14 of september, 2007. The instrument is composed
of 6400 fibers and, has a changeable scale on the sky that was set to
0.33$"$ per fiber, to obtain our data. The image is formed by a matrix
of 40x40 fibers, which gave us a coverage of 13.2$"$x13.2$"$ on the
sky. We obtained observations in high-resolution mode, with a pixel
scale of 0.6~\AA~pix$^{-1}$ with a useable range from 3900~\AA\ to
7000~\AA, considering both the blue and red grisms of the
spectrograph. The object was observed with an exposure time of 300
seconds in both red and blue configurations. The reduction was
performed with the VIMOS pipelines available at the instrument website
\footnote{http://www.eso.org/sci/facilities/paranal/instruments/vimos/}

\section{Results}

\begin{figure}[!ht]
\begin{center}
    \includegraphics[scale=0.32]{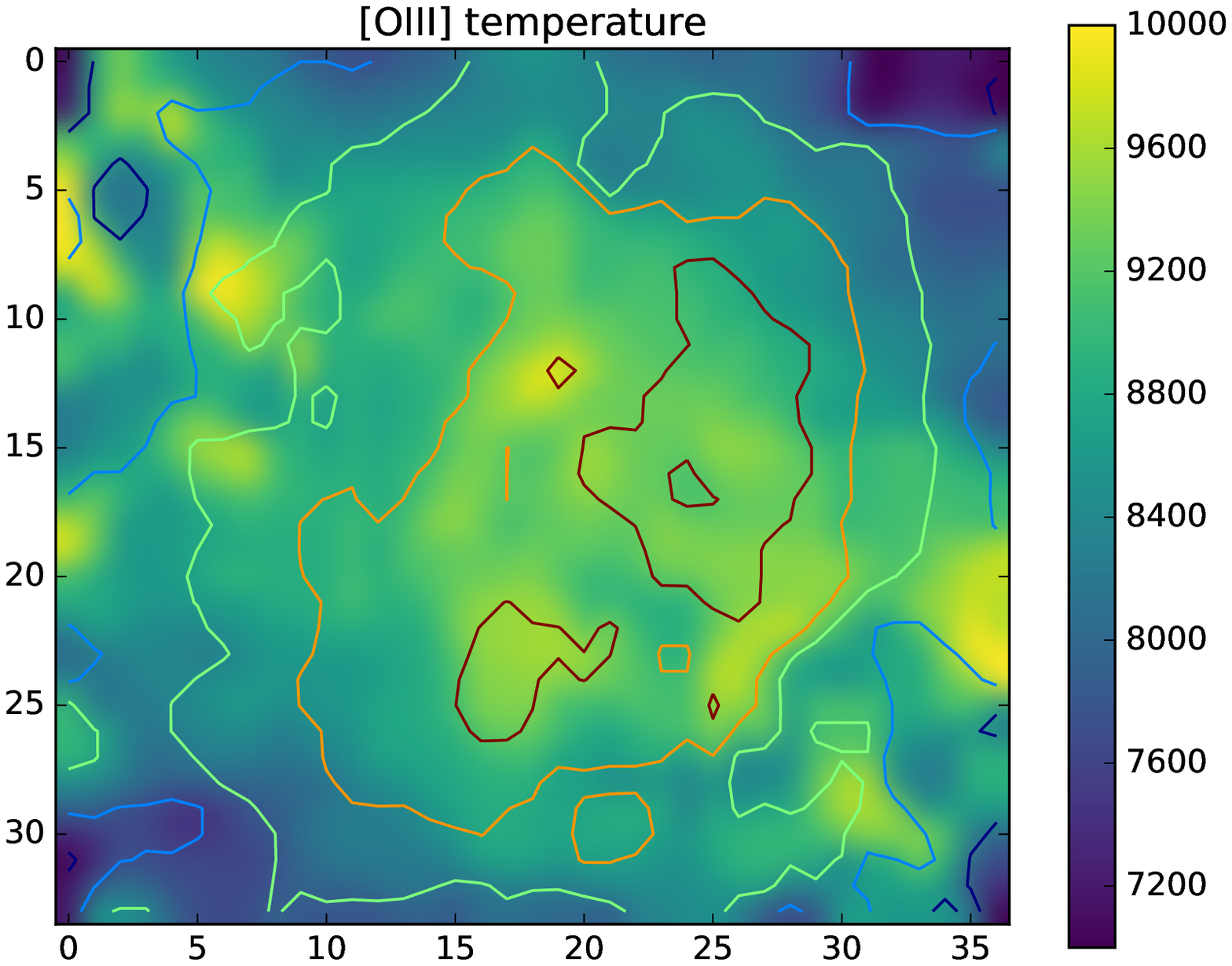}
    \includegraphics[scale=0.32]{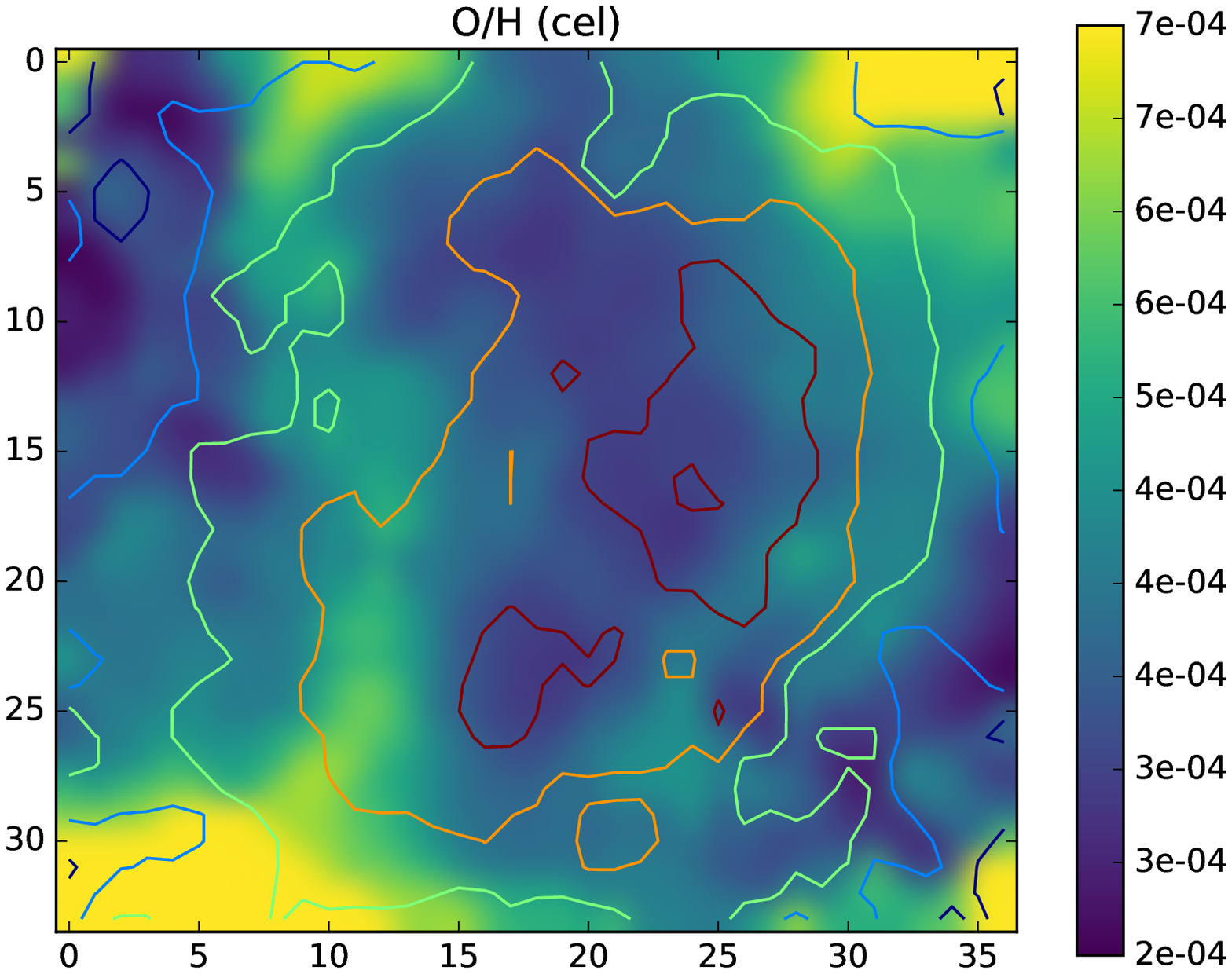}\\
    \includegraphics[scale=0.32]{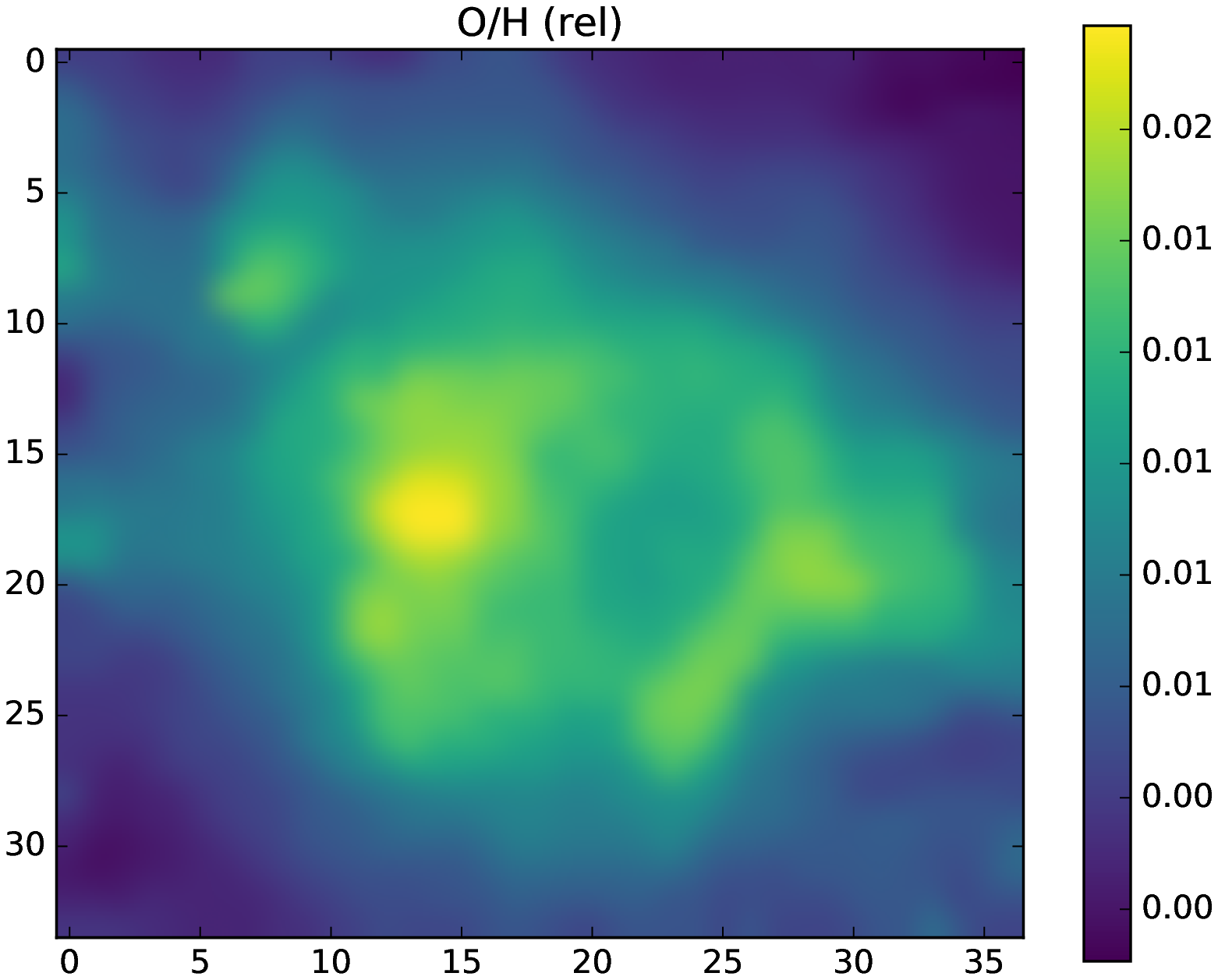}
    \includegraphics[scale=0.32]{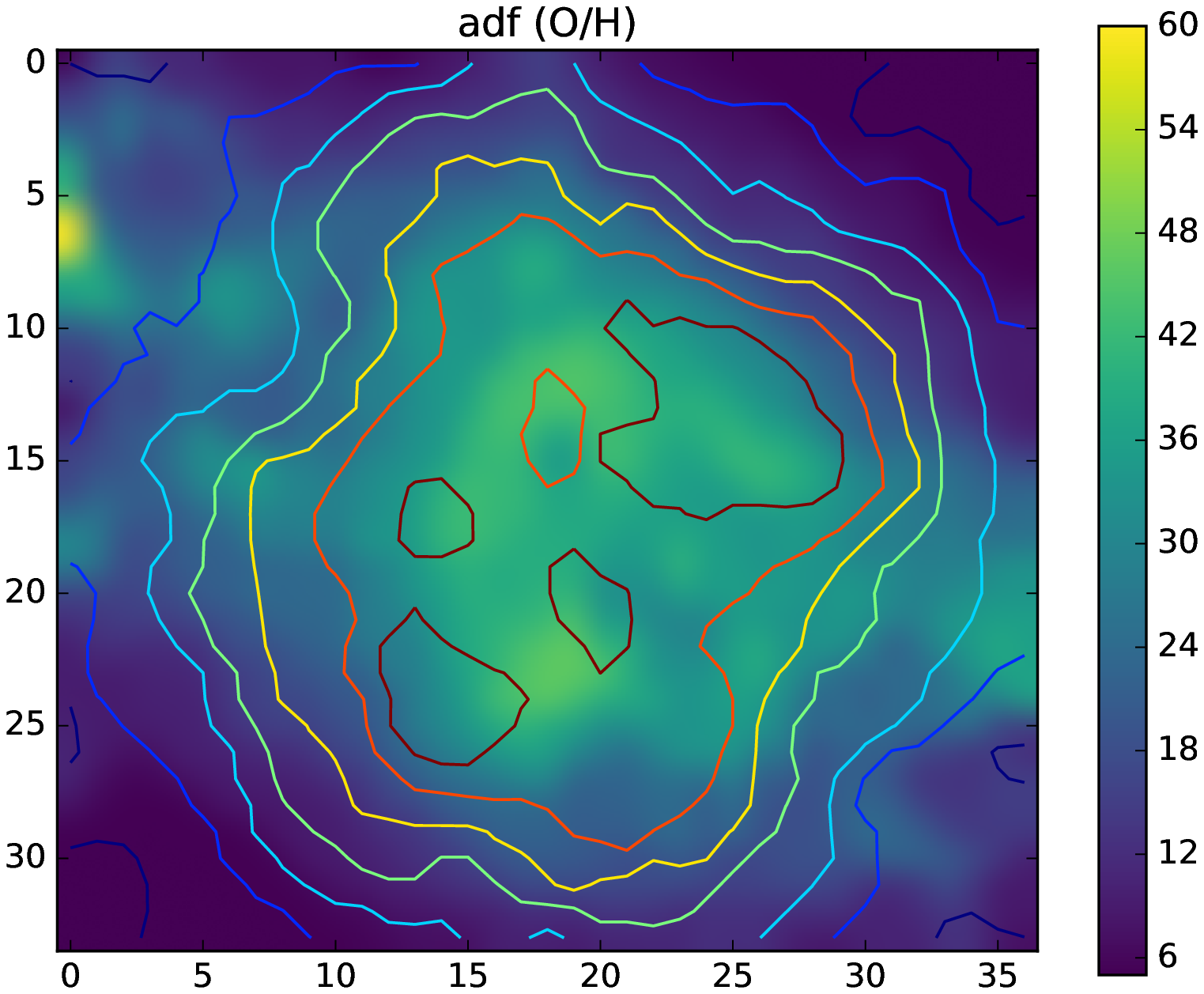}\\
    \caption{Temperature (upper left), abundances obtained from CELs
      (upper right), both with the [OIII]4363 contours overlaid, from
      RELs (lower left) and ADF maps (lower right with the OII 4650
      contours overlaid) for NGC 6778.}
    \label{line-maps}
\end{center}
\end{figure}

In Fig.~\ref{line-maps} we show the emission-line maps for some of the
most important lines we detected, which were also used to obtain
diagnostics of the electron densities and temperatures. 

Because we do not deal with integrated fluxes, the typical
signal-to-noise ratio in a volumetric pixel (voxel) of the data cube
can be significantly lower in comparison to usual long-slit data. In
practice, the trade-off of spatial resolution is lower signal to noise
ratio (S/N) in a given pixel of the observed map. The limited S/N may
result in significant noise when spatially resolved diagnostic ratio
maps are computed.  To improve the quality of the final maps we
applied a median noise filter to remove some of the noise, especially
in the low S/N regions.

To work with the emission line maps presented in Fig.~\ref{line-maps}
and derive the nebular properties (internal extinction, electron
densities, and temperatures and abundances), we used a set of Python
scripts and the NEAT software described in detail in Wesson et
al. (2012).

The code computes the extinction coefficient c(H$\beta$) using the
available Balmer lines. The extinction-corrected emission-line maps
were used to derive the spatial distribution of the electron density
(N$_e$) and temperature (T$_e$) of the nebula as well as abundances
and abundance discrapancy factors (ADFs). For the calculations we
adopted the extinction law of Howarth (1983). The atomic data used for
He I abundances is from Smits (1996). The ICF scheme used to correct
for unseen ions was that of Delgado-Inglada et al. (2014). The results
for the physical and chemical properties are shown in
Figs 4. through 8.

The results obtained here coroborate previous results from long-slit
spectra presented by Jones et al. (2016) and expand on the ones from
García-Rojas et al. (2016).


\section{Conclusion}

The planetary nebula NGC 6778 has been studied with data from the VLT
IFU isntrument from where we collected high resolution spatially
resolved spectra. The nebula shows a density enhanced waist which
correlates with the bright emissions regions seen in both [SII] and
[NII] lines. The temperature estimated from the low ionization [NII]
lines as well as the [OIII] lines, shows hotter regions along the axis
perpendicular to the main waist.

The abundance maps also show structural variations that seem to follow
the axis perpendicular to the main waist. Interestingly the abundance
variations obtained from recombination lines show an increase towards
the central region of the nebula while the abundances obtained from
colisional lines show increases that are coincident with the higher
density waist.

The [OIII]4363 emission resembles the OII recombination emission but
not the strong [OIII]5007 emission which may be due to presence of a
high-density, H-poor gas component in the inner regions of the
nebula. The H-poor component could be the source of the OII emission
and the cause of the abundance discrepancy. The ADF maps show larger
discrepancies correlated with the axis perpendicular to the waist,
consistent with values obtained by Jones et al. (2016). These results
may provide important constraints for the existence and formation
times of jets (Guerrero \& Miranda ,2012). All this information will
also help in constraining the role of binaries and their relation to
the geometry and ejection mechanism of the metal rich (H-poor)
component.

\acknowledgments This research was performed using the computer
cluster {\bf Giskard} and assistance of the Laboratório de Astrofísica
Computacional da Universidade Federal de Itajubá (LAC-UNIFEI). The
LAC-UNIFEI Giskard cluster was acquired by a Capes Pró-Equipamentos
Grant in 2014.

\end{document}